\documentclass[twocolumn,showpacs,preprintnumbers,amsmath,amssymb]{revtex4}

\usepackage{graphicx}
\newcommand{\mueff}{{\mu^\text{eff}}}
\newcommand{\Rmin}{{R_\text{min}}}

\begin{document}

\preprint{journal ref: Phys.\ Rev.\ Lett., {\bf 98}, 018301 (2007)}

\title{Refraction of shear zones in granular materials}

\author{Tam\'as Unger}
\affiliation{%
Research group ``Theory of Condensed Matter'' of the
Hungarian Academy of Sciences,
Institute of Physics, 
Budapest University
of Technology and Economics,
H-1111 Budapest, Hungary}%
\affiliation{Department of Physics, University Duisburg-Essen,
 D-47048 Duisburg, Germany}


\date{\today}

\begin{abstract}
We study strain localization in slow shear flow focusing on layered
granular materials. A heretofore unknown effect is presented here. We show
that shear zones are refracted at material interfaces in analogy with
refraction of light beams in optics. This phenomenon can be
obtained as a consequence of a recent variational model of shear
zones. The predictions of the model are tested and confirmed by 3D discrete
element simulations. We found that shear zones follow Snell's law of light
refraction. 
\end{abstract}



\maketitle

The motion of granular materials, such as sand, is difficult to predict, as
they behave neither like elastic solids nor like normal fluids. When they
yield under stress and start flowing, the material consists of large,
almost solid parts, and the relative motion is confined to narrow regions
between them, called shear zones or shear bands
\cite{Mueth00,Fenistein03,GDRMiDi04}. Shear zones represent material
failure and therefore play a crucial role in geophysics (geological
faults), in engineering (building foundations) and in industrial
processes. It is an important problem of these fields to predict where the
failure takes place. This is a difficult task for those shear zones which
arise in the bulk of the material far from the confining walls
\cite{Fenistein03,Fenistein04,Luding04,Cheng06,Unger04a,Torok06}.

Nature often stratifies granular media, i.e. different materials are
deposited in distinct layers. We study here the question what effect
such inhomogeneities have on shear zones.  Based on a recent theory and
computer simulations we show that shear zones are refracted when they pass
through the interface between different granular layers. Moreover, the
angle of refraction obeys a law analogous to Snell's law of light
refraction which reveals an unexpected analogy between granular media and
geometric optics. The effect 
of refraction presented here influences the position of shear zones in
various kinds of materials including powders, sand, soil and rock layers
providing implications on the behavior of geological faults
\cite{Schultz05,Scott96}.

Recently the variational principle of minimum dissipation was successfully
applied \cite{Unger04a,Torok06} to describe the non-trivial shape of the
shear zone in a homogeneous granular material which is sheared in a
modified Couette cell \cite{Fenistein04,Luding04,Cheng06}. Minimum rate of
energy dissipation is a widely applied selection principle \cite{Jaynes80}
to determine which steady states are realized in nature. When applying this
principle to the cylindrical geometry of the modified Couette cell
\cite{Unger04a}, not only the non-trivial position of the shear zone could
be calculated without any fitting parameter, but new, closed shear zones
were predicted which were discovered also in experiments \cite{Fenistein06}
and simulations \cite{Cheng06}. In the following, a so far unknown but
measurable effect will be derived from the same principle.
It concerns the shape of shear zones in layered granular materials.

\begin{figure}
\includegraphics[scale=0.4]{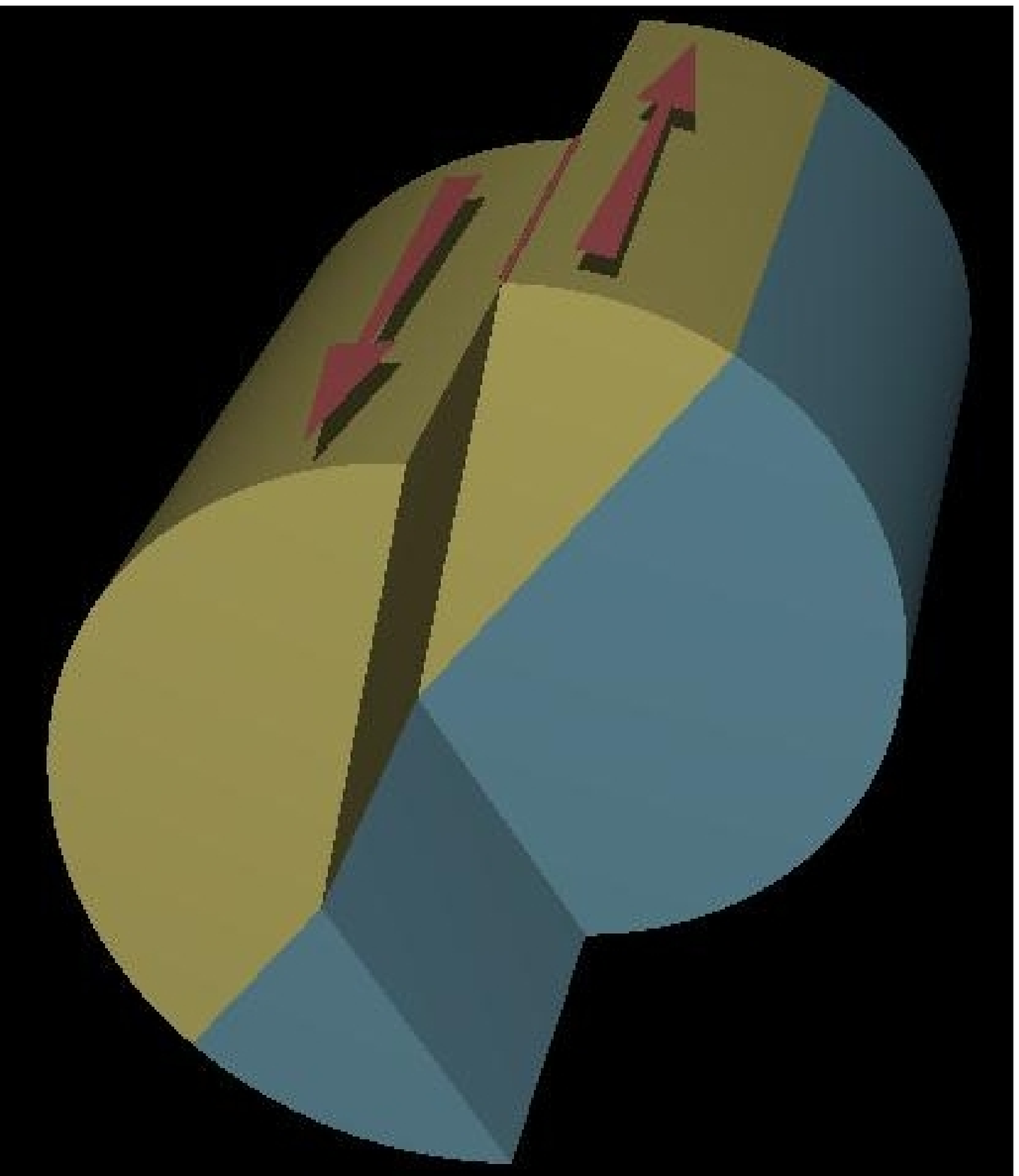}
\includegraphics[scale=0.2]{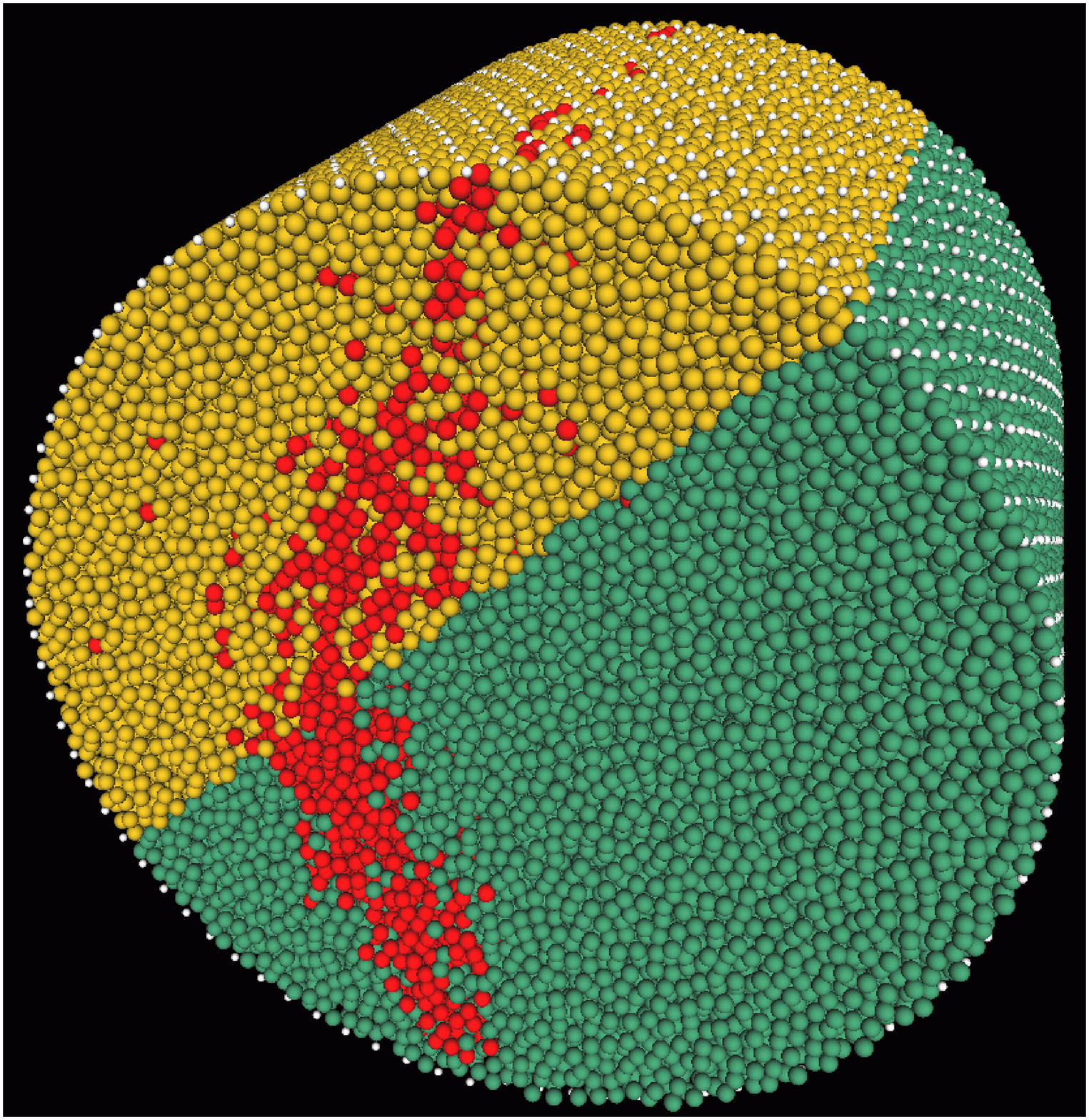}
\includegraphics[scale=0.62]{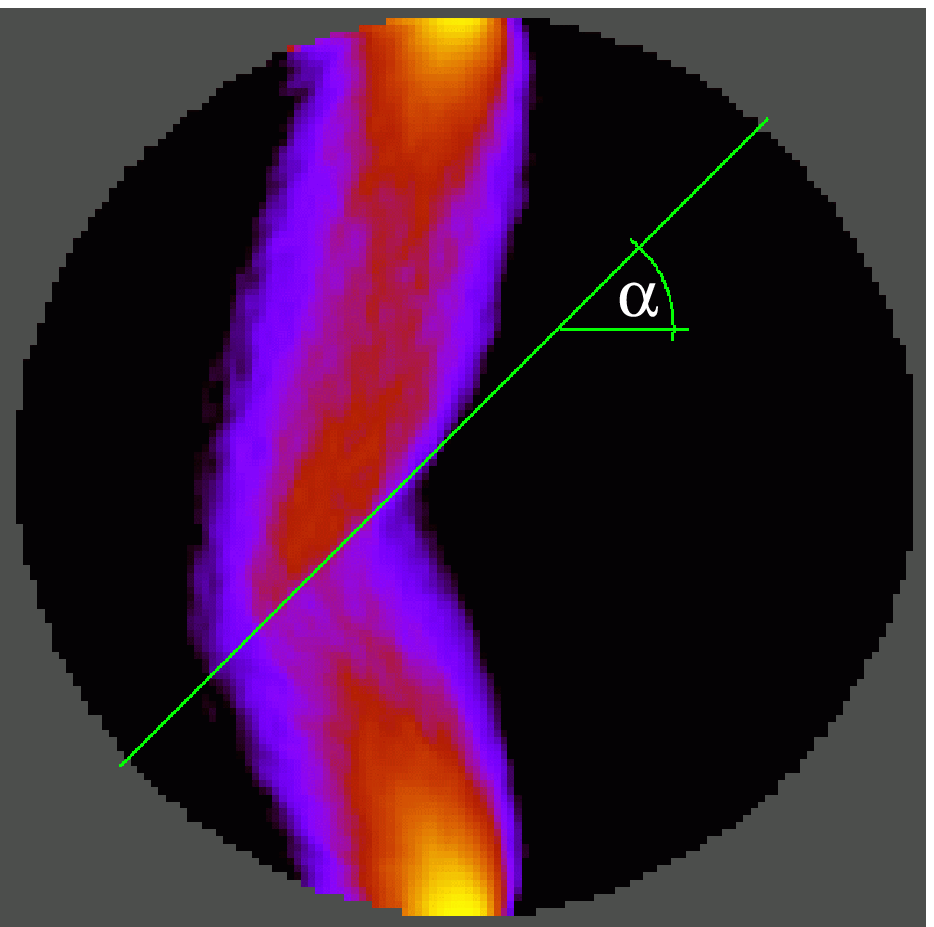}
\caption{\label{fig:refraction} 
  (color online) a, Schematic view showing the shear cell and the
  refraction of the shear zone. Dense granular material is sheared between
  two half-cylinders. The shear direction is indicated by the arrows. b, A
  snapshot of the simulation. The upper light-gray (yellow) material and
  lower dark-gray (blue) material have different frictional properties
  (blue beads have stronger friction). If the velocity of any bead drops
  below half of the external shear velocity, it is overpainted by dark-gray
  (red). Despite the fluctuations of single bead-velocities, a dark (red)
  zone appears in the bulk that separates the two moving parts of the
  system. c, The spatial distribution of the shear rate which shows the
  structure of the shear zone. Lighter colors represent stronger
  deformations (orange means the largest shear rate, followed by red and
  blue, while black means almost no deformation). The straight line
  indicates the material interface.}
\end{figure}

This phenomenon can be understood best, if the side effects of gravity and
curved shearing are absent. Therefore a long cylindrical container is
considered, which is cut along its axis into two halves
(Fig.~\ref{fig:refraction}a). (In computer simulations periodic boundary
conditions in the axial direction are applied.) The container is completely
filled with two granular materials having different frictional
properties. They are separated by a planar interface parallel to the axis,
but at an angle to the cut of the container. The system is sheared
quasi-statically \cite{GDRMiDi04} by moving the two halfs of the cylinder
wall slowly parallel to the axis in opposite directions. This creates a
shear zone starting and ending where the container wall is cut. In order to
find the shape
of the zone in between we apply the principle of minimum dissipation. This, in
zero width approximation \cite{Unger04a}, leads to the following
variational problem (explained below): 
\begin{equation}
\int v p \mueff \text{d}S = \text{min.}
\end{equation}
Here the local rate of energy dissipation is integrated over the whole
surface of the shear zone. The shear zone is regarded as being infinitely
thin, outside the shear zone no deformation and no dissipation takes
place. The local dissipation rate per unit area $ v p \mueff$ is obtained
by the sliding velocity $v$ between the two sides of the shear zone times
the shear stress $p \mueff$. The parameters $p$ and $\mueff$ denote the
overall pressure in the system and the coarse grained effective friction
coefficient \cite{GDRMiDi04}. The effective friction has different values
for the two materials while $v$ and $p$ are taken constant for our setup
\footnote{By contrast, $p$ and $v$ depend strongly on the position in the
  modified Couette cell. This is caused by gravity (for $p$) and by curved
  shearing (for $v$) \cite{Unger04a}. For clarity, these effects are
  avoided in our case.}.  
The surface that minimizes the total rate of dissipation provides the shape
of the shear zone. This means that the system yields along the surface
where it has the least resistance against the external shear.

The system is translation invariant along the axis, therefore the above
surface integral is reduced to a line integral. We have to find the path
(cross section of the shear zone) which connects two fixed points and along which the integral of $\mueff$ is
minimal. This problem has exactly the form of Fermat's principle of
geometric optics where the effective friction plays the role of the index
of refraction. It is known that Fermat's principle of the shortest
travel time leads to refraction of light beams therefore one can expect a
similar phenomenon of refraction also for shear zones.

In order to check this prediction we carried out computer simulations of
$10^5$ frictional and hard spherical beads
(Fig.~\ref{fig:refraction}b). Based on a standard discrete element method
(DEM) \cite{Jean99,Brendel04} the motion of each bead is determined by
Newton's equation. This grain-level dynamics leads to a collective steady
state flow where we measure the coarse-grained velocity field. The
coarse-grained velocities are parallel to the cylinder axis. The local
strain rate, which is calculated from the velocity field, corresponds to
pure shear deformation (without volumetric strain). The spatial
distribution of the magnitude of the shear rate is shown in
Fig.~\ref{fig:refraction}c. As predicted, the arising shear zone departs
from the cut plane and its direction is changed significantly at the
material interface (Fig.~\ref{fig:refraction}b, c). Note, that the shear
zone we obtained is quite wide relative to the system size. The relative
width becomes smaller for larger systems \cite{Fenistein04} which, however,
would demand also a computational effort out of reach. Still, the effect of
refraction is clearly shown.

The analogy with light refraction provides quantitative predictions that
can be measured in numerical and real experiments. Fermat's principle
indicates that shear zones have to follow Snell's law of light refraction
which will be tested here by DEM simulations. In order to do so we
characterize the behavior of the shear zone by two angles $\phi_1$ and
$\phi_2$ which are analogous to the incident and refractive angles used in
geometric optics. These are defined with the help of the shear deformation
(Fig.~\ref{fig:refraction}c) as follows: first, the refraction point is
identified as the point along the interface where the local shear rate has
its maximum. Then the refraction point is connected to the starting and
ending points of the shear zone by straight lines. Thus the direction of
the shear zone on each side of the interface is determined. These
directions provide the angles $\phi_1$ and $\phi_2$ with respect to the
normal of the interface. Another angle, which is an important input
parameter of the shear test, is the tilt angle of the interface $\alpha$
between the normal of the cut plane and the interface
(Fig.~\ref{fig:refraction}c). For various values of $\alpha$ we measured
$\phi_1$ and $\phi_2$. If the interface is perpendicular to the cut plane
($\alpha=0$) no refraction was observed, however, as $\alpha$ is increased
the refraction of the shear zone becomes more and more pronounced.

If we apply Snell¢s law to our case it indicates that, no matter how the
angles $\phi_1$ and $\phi_2$ are changed, the ratio of $\sin{\phi_1}$ and
$\sin{\phi_2}$ has to be independent of the tilt angle $\alpha$.
Furthermore it is expected that this ratio can be expressed by the
effective frictions:
\begin{equation}
\frac{\sin{\phi_1}}{\sin{\phi_2}} = \frac{\mueff_2}{\mueff_1}
\end{equation}

First we report the influence of $\alpha$ observed in simulations. We
used materials that contained hard spheres of radii distributed uniformly
between $\Rmin$ and $1.3 \, \Rmin$. This polydispersity was needed to avoid
shear induced crystallization \cite{Tsai03}. In order to control the
frictional properties of the materials we varied the value of the
microscopic friction $\mu$ which is the friction coefficient at the
particle-particle contacts. It is important to note that the microscopic
friction $\mu$ and the effective friction $\mueff$ are not the same
although they are closely related as we will see later. We simulated three
systems called $A$, $B$ and $C$. The total number of particles was $100000$
for system $A$ and $50000$ for systems $B$ and $C$. For each case the shear
cell was filled with two materials with microscopic frictions $\mu_1$ and
$\mu_2$. For systems $A$ and $B$ $\mu_1=0$ and $\mu_2=0.5$ while for system
$C$ $\mu_1=0.1$ and $\mu_2=0.5$. For all systems the radius of the
container was approximately $65 \, \Rmin$.

\begin{figure}
\includegraphics[scale=0.65]{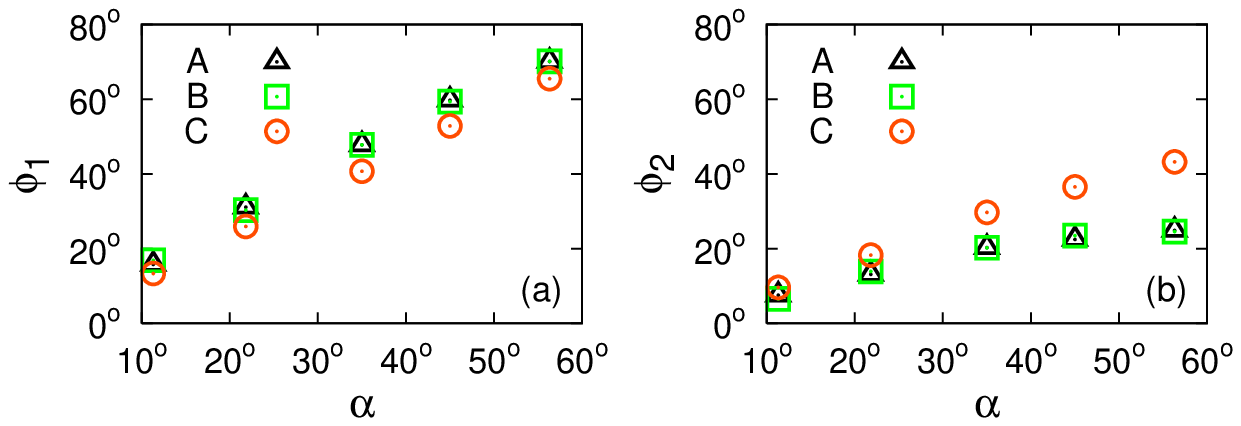}
\includegraphics[scale=0.65]{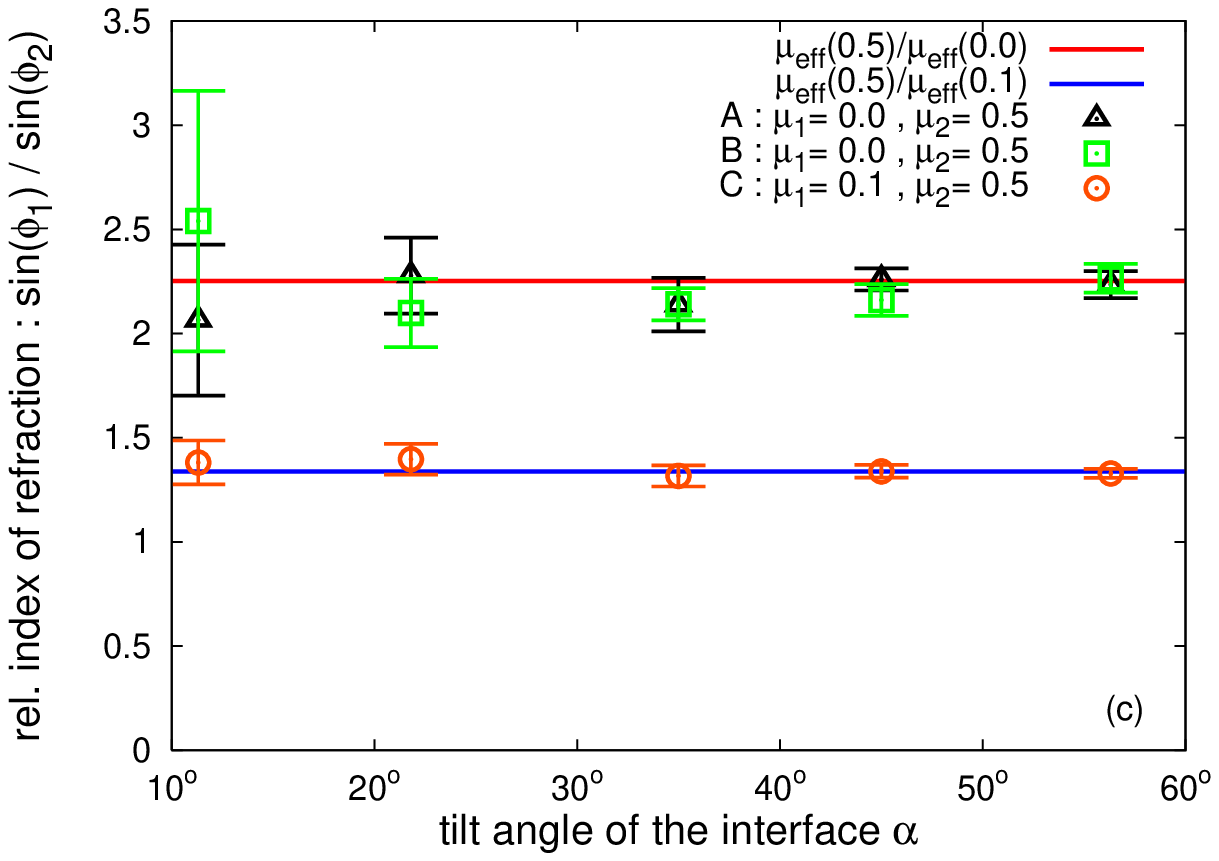}
\caption{\label{fig:Snellslaw} (color online) $\phi_1$ (a), $\phi_2$ (b)
  and the relative index of refraction (c) measured for shear zones in
  computer simulations. The horizontal axis show the tilt angle of the
  interface between the two materials. Data points
  denoted by triangle, square and circle stand for systems A, B and C,
  respectively. The pairs of microscopic friction that are used in these
  systems are indicated in the figure. The straight lines show the
  theoretically predicted values (the upper line for system A and B, the
  lower one for system C).}
\end{figure}

The influence of the tilt angle of the interface is presented in
Fig.~\ref{fig:Snellslaw}. The computer simulations show that $\phi_1$ and
$\phi_2$ vary strongly but $\sin{\phi_1} / \sin{\phi_2}$ remains constant
for a wide range of the angle $\alpha$. This ratio seems to depend only on
the materials in which the shear zone was created, in full agreement with
the theoretical considerations.

Next we deal with the role of the effective friction. It is known that if a
dense noncohesive granular material is sheared then after a short transient
\cite{Craig04} it reaches a well defined resistance against shear
deformation provided the flow is slow enough (quasistatic shear). Thus a
material parameter, the effective friction coefficient, can be defined
which turns out to be independent of the preparation history, the actual
shear rate and the confining pressure \cite{GDRMiDi04,Radjai04}. The effective
friction $\mueff$ is given by the shear stress divided by the normal
stress, both measured in the plane along which the shear deformation takes
place.

\begin{figure}
\includegraphics[scale=0.65]{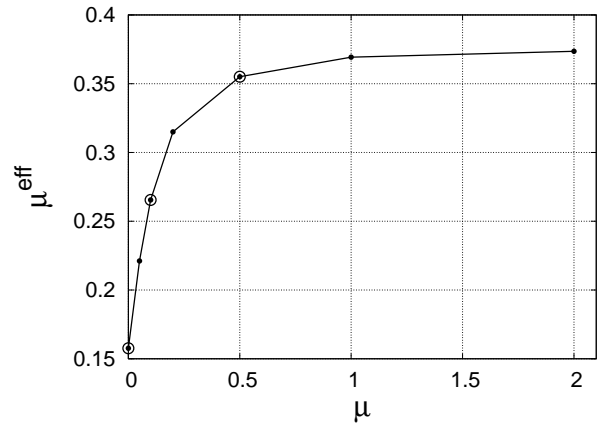}
\caption{\label{fig:mu-mueff} Effective friction versus microscopic
friction. The open circles correspond to the materials used in systems $A$,
$B$ and $C$.}
\end{figure}

The quantity $\mueff$ is not a microscopic input parameter of the computer
simulation, thus the question arises what values $\mueff$ has for the
materials that appeared in systems $A$, $B$ and $C$. This can be deduced
from simulations which are independent of the previous refraction tests: we
put the same materials into a 3D rectangular box under plane shear
\cite{GDRMiDi04} measure the components of the stress tensor
\cite{Christoffersen81} and calculate the value of $\mueff$. In this way we
achieve a calibration curve that provides a one to one correspondence
between microscopic and effective friction, see Fig.~\ref{fig:mu-mueff}.

With that we arrived to a point where we can make quantitative statements
about the extent of the refraction based on Snell's law. E.g.\ for system
$A$, with help of the prescribed values of $\mu_1$ and $\mu_2$, the ratio
$\mueff_2 / \mueff_1$ can be calculated and compared to the data
$\sin{\phi_1} / \sin{\phi_2}$ recorded in the refraction test. This is done
in Fig.~\ref{fig:Snellslaw} for systems $A$, $B$ and $C$.
A surprisingly good agreement is found between the predicted and measured
values which holds for all systems and for various tilt angles of the
interface. This is our main result.

Dealing with shear localization in layered granular materials we
investigated the predictions of a recent variational model and compared
them to computer simulations. The results of the present work convey two
messages. First, we gave verification of the variational approach to
granular shear flows. We tested the model in a new situation and, according
to the numerical data, it gave an excellent description of the behavior of
the shear zones. Second, an interesting
analogy between granular flow and geometric optics is revealed. We showed
that shear zones are refracted at material interfaces similarly to light
beams. The phenomenon presented here should be accessible by experiments.

We are grateful to J. Kert\'esz and D.E. Wolf for their support and help.
Support by grant OTKA T049403 and by the G.I.F. research grant
I-795-166.10/2003 is acknowledged.

\bibliography{granu}

\begin{thebibliography}{18}
\expandafter\ifx\csname natexlab\endcsname\relax\def\natexlab#1{#1}\fi
\expandafter\ifx\csname bibnamefont\endcsname\relax
  \def\bibnamefont#1{#1}\fi
\expandafter\ifx\csname bibfnamefont\endcsname\relax
  \def\bibfnamefont#1{#1}\fi
\expandafter\ifx\csname citenamefont\endcsname\relax
  \def\citenamefont#1{#1}\fi
\expandafter\ifx\csname url\endcsname\relax
  \def\url#1{\texttt{#1}}\fi
\expandafter\ifx\csname urlprefix\endcsname\relax\def\urlprefix{URL }\fi
\providecommand{\bibinfo}[2]{#2}
\providecommand{\eprint}[2][]{\url{#2}}

\bibitem[{\citenamefont{Mueth et~al.}(2000)\citenamefont{Mueth, Debregeas,
  Karczmar, Eng, Nagel, and Jaeger}}]{Mueth00}
\bibinfo{author}{\bibfnamefont{D.~M.} \bibnamefont{Mueth}},
  \bibinfo{author}{\bibfnamefont{G.~F.} \bibnamefont{Debregeas}},
  \bibinfo{author}{\bibfnamefont{G.~S.} \bibnamefont{Karczmar}},
  \bibinfo{author}{\bibfnamefont{P.~J.} \bibnamefont{Eng}},
  \bibinfo{author}{\bibfnamefont{S.~R.} \bibnamefont{Nagel}}, \bibnamefont{and}
  \bibinfo{author}{\bibfnamefont{H.~M.} \bibnamefont{Jaeger}},
  \bibinfo{journal}{Nature} \textbf{\bibinfo{volume}{406}},
  \bibinfo{pages}{385} (\bibinfo{year}{2000}).

\bibitem[{\citenamefont{Fenistein and van Hecke}(2003)}]{Fenistein03}
\bibinfo{author}{\bibfnamefont{D.}~\bibnamefont{Fenistein}} \bibnamefont{and}
  \bibinfo{author}{\bibfnamefont{M.}~\bibnamefont{van Hecke}},
  \bibinfo{journal}{Nature} \textbf{\bibinfo{volume}{425}},
  \bibinfo{pages}{256} (\bibinfo{year}{2003}).

\bibitem[{\citenamefont{GDR\_MiDi}(2004)}]{GDRMiDi04}
\bibinfo{author}{\bibnamefont{GDR\_MiDi}}, \bibinfo{journal}{Eur. Phys. J. E}
  \textbf{\bibinfo{volume}{14}}, \bibinfo{pages}{341} (\bibinfo{year}{2004}).

\bibitem[{\citenamefont{Fenistein et~al.}(2004)\citenamefont{Fenistein, van~de
  Meent, and van Hecke}}]{Fenistein04}
\bibinfo{author}{\bibfnamefont{D.}~\bibnamefont{Fenistein}},
  \bibinfo{author}{\bibfnamefont{J.~W.} \bibnamefont{van~de Meent}},
  \bibnamefont{and} \bibinfo{author}{\bibfnamefont{M.}~\bibnamefont{van
  Hecke}}, \bibinfo{journal}{Phys. Rev. Lett.} \textbf{\bibinfo{volume}{92}},
  \bibinfo{pages}{094301} (\bibinfo{year}{2004}).

\bibitem[{\citenamefont{Luding}(2004)}]{Luding04}
\bibinfo{author}{\bibfnamefont{S.}~\bibnamefont{Luding}}, in
  \emph{\bibinfo{booktitle}{The Physics of Granular Media}}
  (\bibinfo{publisher}{Wiley-VCH}, \bibinfo{address}{Weinheim},
  \bibinfo{year}{2004}), pp. \bibinfo{pages}{299--324}.

\bibitem[{\citenamefont{Cheng et~al.}(2006)\citenamefont{Cheng, Lechman,
  Fernandez-Barbero, Grest, Jaeger, Karczmar, M\"obius, and Nagel}}]{Cheng06}
\bibinfo{author}{\bibfnamefont{X.}~\bibnamefont{Cheng}},
  \bibinfo{author}{\bibfnamefont{J.~B.} \bibnamefont{Lechman}},
  \bibinfo{author}{\bibfnamefont{A.}~\bibnamefont{Fernandez-Barbero}},
  \bibinfo{author}{\bibfnamefont{G.~S.} \bibnamefont{Grest}},
  \bibinfo{author}{\bibfnamefont{H.~M.} \bibnamefont{Jaeger}},
  \bibinfo{author}{\bibfnamefont{G.~S.} \bibnamefont{Karczmar}},
  \bibinfo{author}{\bibfnamefont{M.~E.} \bibnamefont{M\"obius}},
  \bibnamefont{and} \bibinfo{author}{\bibfnamefont{S.~R.} \bibnamefont{Nagel}},
  \bibinfo{journal}{Phys. Rev. Lett.} \textbf{\bibinfo{volume}{96}},
  \bibinfo{pages}{038001} (\bibinfo{year}{2006}).

\bibitem[{\citenamefont{Unger et~al.}(2004)\citenamefont{Unger, T\"or\"ok,
  Kert\'esz, and Wolf}}]{Unger04a}
\bibinfo{author}{\bibfnamefont{T.}~\bibnamefont{Unger}},
  \bibinfo{author}{\bibfnamefont{J.}~\bibnamefont{T\"or\"ok}},
  \bibinfo{author}{\bibfnamefont{J.}~\bibnamefont{Kert\'esz}},
  \bibnamefont{and} \bibinfo{author}{\bibfnamefont{D.~E.} \bibnamefont{Wolf}},
  \bibinfo{journal}{Phys. Rev. Lett.} \textbf{\bibinfo{volume}{92}},
  \bibinfo{pages}{214301} (\bibinfo{year}{2004}),
  \bibinfo{note}{cond-mat/0401143}.

\bibitem[{\citenamefont{T\"or\"ok et~al.}()\citenamefont{T\"or\"ok, Unger,
  Kert\'esz, and Wolf}}]{Torok06}
\bibinfo{author}{\bibfnamefont{J.}~\bibnamefont{T\"or\"ok}},
  \bibinfo{author}{\bibfnamefont{T.}~\bibnamefont{Unger}},
  \bibinfo{author}{\bibfnamefont{J.}~\bibnamefont{Kert\'esz}},
  \bibnamefont{and} \bibinfo{author}{\bibfnamefont{D.~E.} \bibnamefont{Wolf}},
  \bibinfo{note}{cond-mat/0607162 (http://www.arxiv.org/abs/cond-mat/0607162),
  accepted in PRE}.

\bibitem[{\citenamefont{Schultz and Siddharthan}(2005)}]{Schultz05}
\bibinfo{author}{\bibfnamefont{R.~A.} \bibnamefont{Schultz}} \bibnamefont{and}
  \bibinfo{author}{\bibfnamefont{R.}~\bibnamefont{Siddharthan}},
  \bibinfo{journal}{Tectonophysics} \textbf{\bibinfo{volume}{411}},
  \bibinfo{pages}{1} (\bibinfo{year}{2005}).

\bibitem[{\citenamefont{Scott}(1996)}]{Scott96}
\bibinfo{author}{\bibfnamefont{D.~R.} \bibnamefont{Scott}},
  \bibinfo{journal}{Nature} \textbf{\bibinfo{volume}{381}},
  \bibinfo{pages}{592} (\bibinfo{year}{1996}).

\bibitem[{\citenamefont{Jaynes}(1980)}]{Jaynes80}
\bibinfo{author}{\bibfnamefont{E.~T.} \bibnamefont{Jaynes}},
  \bibinfo{journal}{Ann. Rev. Phys. Chem.} \textbf{\bibinfo{volume}{31}},
  \bibinfo{pages}{579} (\bibinfo{year}{1980}).

\bibitem[{\citenamefont{Fenistein et~al.}(2006)\citenamefont{Fenistein, van~de
  Meent, and van Hecke}}]{Fenistein06}
\bibinfo{author}{\bibfnamefont{D.}~\bibnamefont{Fenistein}},
  \bibinfo{author}{\bibfnamefont{J.-W.} \bibnamefont{van~de Meent}},
  \bibnamefont{and} \bibinfo{author}{\bibfnamefont{M.}~\bibnamefont{van
  Hecke}}, \bibinfo{journal}{Phys. Rev. Lett.} \textbf{\bibinfo{volume}{96}},
  \bibinfo{pages}{118001} (\bibinfo{year}{2006}).

\bibitem[{\citenamefont{Jean}(1999)}]{Jean99}
\bibinfo{author}{\bibfnamefont{M.}~\bibnamefont{Jean}},
  \bibinfo{journal}{Comput. Methods Appl. Mech. Engrg.}
  \textbf{\bibinfo{volume}{177}}, \bibinfo{pages}{235} (\bibinfo{year}{1999}).

\bibitem[{\citenamefont{Brendel et~al.}(2004)\citenamefont{Brendel, Unger, and
  Wolf}}]{Brendel04}
\bibinfo{author}{\bibfnamefont{L.}~\bibnamefont{Brendel}},
  \bibinfo{author}{\bibfnamefont{T.}~\bibnamefont{Unger}}, \bibnamefont{and}
  \bibinfo{author}{\bibfnamefont{D.~E.} \bibnamefont{Wolf}}, in
  \emph{\bibinfo{booktitle}{The Physics of Granular Media}}
  (\bibinfo{publisher}{Wiley-VCH}, \bibinfo{address}{Weinheim},
  \bibinfo{year}{2004}), pp. \bibinfo{pages}{325--343}.

\bibitem[{\citenamefont{Tsai et~al.}(2003)\citenamefont{Tsai, Voth, and
  Gollub}}]{Tsai03}
\bibinfo{author}{\bibfnamefont{J.~C.} \bibnamefont{Tsai}},
  \bibinfo{author}{\bibfnamefont{G.~A.} \bibnamefont{Voth}}, \bibnamefont{and}
  \bibinfo{author}{\bibfnamefont{J.~P.} \bibnamefont{Gollub}},
  \bibinfo{journal}{Phys. Rev. Lett.} \textbf{\bibinfo{volume}{91}},
  \bibinfo{pages}{064301} (\bibinfo{year}{2003}).

\bibitem[{\citenamefont{Craig}(2004)}]{Craig04}
\bibinfo{author}{\bibfnamefont{R.~F.} \bibnamefont{Craig}},
  \emph{\bibinfo{title}{Craig's soil mechanics}} (\bibinfo{publisher}{Spon
  Press}, \bibinfo{address}{New York}, \bibinfo{year}{2004}).

\bibitem[{\citenamefont{Radjai and Roux}(2004)}]{Radjai04}
\bibinfo{author}{\bibfnamefont{F.}~\bibnamefont{Radjai}} \bibnamefont{and}
  \bibinfo{author}{\bibfnamefont{S.}~\bibnamefont{Roux}}, in
  \emph{\bibinfo{booktitle}{The Physics of Granular Media}}
  (\bibinfo{publisher}{Wiley-VCH}, \bibinfo{address}{Weinheim},
  \bibinfo{year}{2004}), pp. \bibinfo{pages}{165--187}.

\bibitem[{\citenamefont{Christoffersen
  et~al.}(1981)\citenamefont{Christoffersen, Mehrabadi, and
  Nemat-Nasser}}]{Christoffersen81}
\bibinfo{author}{\bibfnamefont{J.}~\bibnamefont{Christoffersen}},
  \bibinfo{author}{\bibfnamefont{M.~M.} \bibnamefont{Mehrabadi}},
  \bibnamefont{and}
  \bibinfo{author}{\bibfnamefont{S.}~\bibnamefont{Nemat-Nasser}},
  \bibinfo{journal}{J. Appl. Mech.} \textbf{\bibinfo{volume}{48}},
  \bibinfo{pages}{339} (\bibinfo{year}{1981}).

\end{thebibliography}

\end{document}